\documentclass[%
 english,
 reprint,
 amsmath,amssymb,
 aps,
]{revtex4-1}
 
\usepackage[T1]{fontenc}
\usepackage[utf8]{inputenc}
\usepackage{babel}
\usepackage{graphicx}

\usepackage{color}
\definecolor{cs1}{RGB}{227, 115, 2}
\definecolor{cs2}{RGB}{6, 137, 179}
\usepackage[unicode=true,
 bookmarks=true,bookmarksnumbered=false,bookmarksopen=false,
 pdfborder={0 0 1},backref=false,colorlinks=true,breaklinks=false]
 {hyperref}
\hypersetup{pdftitle={Fluctuational electrodynamics for nonlinear materials in and out
 of thermal equilibrium},
 pdfauthor={Heino Soo and Matthias Krüger},
 linkcolor=cs2, citecolor=cs1}

\begin{document}

\title{Fluctuational electrodynamics for nonlinear materials in and out of thermal equilibrium}

\author{Heino Soo}
\affiliation{4th Institute for Theoretical Physics, Universität Stuttgart, Germany}
\affiliation{Max Planck Institute for Intelligent Systems, 70569 Stuttgart, Germany}
\author{Matthias Krüger}
\affiliation{4th Institute for Theoretical Physics, Universität Stuttgart, Germany}
\affiliation{Max Planck Institute for Intelligent Systems, 70569 Stuttgart, Germany}

\begin{abstract}
We develop fluctuational electrodynamics for media with nonlinear optical response in and out of thermal equilibrium. Starting from the stochastic nonlinear Helmholtz equation and using the fluctuation dissipation theorem, we obtain perturbatively a deterministic nonlinear Helmholtz equation for the average field, the physical linear response, as well as the fluctuations and Rytov currents. We show that the effects of nonlinear optics, in or out of thermal equilibrium, can be taken into account with an effective, system-aware dielectric function. We discuss the heat radiation of a planar, nonlinear surface, showing that Kirchhoff's must be applied carefully. We find that the spectral emissivity of a nonlinear nanosphere can in principle be negative, implying the possibility of heat flow reversal for specific frequencies. 
\end{abstract}

\maketitle

\section{Introduction}

Fluctuational electrodynamics (FE) has been instrumental in describing
physical phenomena which involve electromagnetic noise, such as the
Casimir effect and heat transfer \cite{Casimir1948a,Lifshitz56,Rytov3,Polder71,Milonni,Bordag}.
By using the fluctuation dissipation theorem (FDT), FE connects the
properties of the involved objects (such as reflection
coefficients) with the fluctuations of charges and fields in or out of equilibrium. This has led to a wealth of
theoretical results which have been tested extensively in experiments
\cite{Lamoreaux1997,Mohideen1998,Bressi2002,Kittel2005,obrecht2007,Rousseau2009, Shen2009,Ottens2011,Kim2015}.
The FE theory has historically focused on systems which respond purely
linearly to the electric field. This simplifies the already formidable
problem, because fluctuations can be treated separately from other
parts of the field due to superposition principle. As a result, however,
many interesting phenomena classified under nonlinear optics have
not been taken into account. These include for example frequency mixing,
the optical Kerr effect, as well as Raman and Brillouin effects \cite{Schubert1986,Boyd2008}.

The concept of fluctuations in systems with a nonlinear response has
been investigated for more than 50 years \cite{Kampen1965,stratonovich2012nonlinear}.
However, research regarding fluctuation phenomena for optically nonlinear
systems appears limited. The noise polarization has been discussed in 
the context of nonlinear macroscopic
quantum electrodynamics \cite{Drummond1990,Scheel2006}.
Equilibrium Casimir forces have been studied from a field theoretical perspective
\cite{kheirandish2011finite}, focusing on the situation of a nonlinear material immersed between two bodies, while Van der Waals forces for objects with nonlinear polarizability were analyzed in Refs.~\cite{Kysylychyn2013,Makhnovets2016,Soo2016c}. A generalization of the framework of FE  to nonlinear materials was performed in Ref.~\cite{Soo2016a}, where it was found that proximity between nonlinear objects can change their effective linear
properties and gives rise to a qualitatively different Casimir force. Also, nonequilibrium cases have been studied, such as heat radiation of a
single nonlinear optical cavity using classical Langevin equations \cite{Khandekar2014,Khandekar2017}. 
However, there seems to be no literature available on theoretical approaches to out
of equilibrium phenomena in nonlinear optics, which are based on the
vector Helmholtz equation. 

In this paper, we extend the framework of Ref.~\cite{Soo2016a}, first providing a more extensive discussion and derivation of the theory for equilibrium processes. We then develop a framework for out of equilibrium scenarios. 
Starting from the nonlinear stochastic Helmholtz equation, we derive the fluctuations of the electromagnetic field and the corresponding Rytov source fluctuations, which are in agreement with the FDT. Based on a local equilibrium assumption, we then derive the fluctuations for bodies at different temperatures and the corresponding heat radiation and transfer formulas.  

We find that Casimir forces as well as heat radiation and transfer can be rationalized in terms of an effective dielectric function (i.e., an effective linear response function), which is to replace the dielectric function in well known formulas for linear materials. We determine the properties of the effective dielectric function and demonstrate that it depends on the shape of an object and the position of other objects. For nonequilibrium scenarios, the dielectric function also depends on the \emph{temperatures} of objects. We carefully discuss in which scenarios the effective dielectric function can be computed and in which cases it contains a divergence from the Green's function at coinciding points. 

Regarding the heat radiation of a body with nonlinear dielectric properties, we discuss the applicability of Kirchoff's law of radiation as well as the (im)possibility to exceed the black body limit of a planar surface. Regarding the radiation of a nonlinear nanosphere,  we show that, in principle, the heat can flow from the colder sphere to a warmer environment in some frequency range.

The manuscript is organized as follows. In Section \ref{sec:Nonlinear-luctuational-electrody}
we introduce the stochastic nonlinear Helmholtz equation. We determine
the equilibrium fluctuations as well as the effective dielectric function
and calculate it numerically for single and parallel plate geometries.
In Section \ref{sec:Thermal-inequilibrium} we 
allow objects to have different temperatures and derive formulas for heat radiation and transfer, which are exemplified for the case of a plate and a nanosphere.


\section{Nonlinear fluctuational electrodynamics in equilibrium\label{sec:Nonlinear-luctuational-electrody}}

\subsection{Stochastic nonlinear Helmholtz equation}

The macroscopic Maxwell's equations describe the
dynamics of the electric ($\mathbf{E}$) and magnetic ($\mathbf{B}$)
fields in matter via the polarization ($\mathbf{P}$)
and magnetization ($\mathbf{M}$) fields. In this article we consider
nonmagnetic materials ($\mathbf{M}=0$). The Maxwell's equations can
then be cast in the form of the well-known Helmholtz equation (also
known as the wave equation), which in frequency space is given as
\begin{equation}
\nabla\times\nabla\times\mathbf{E}-\frac{\omega^{2}}{c^{2}}\mathbf{E}-\frac{1}{\varepsilon_{0}}\frac{\omega^{2}}{c^{2}}\mathbf{P}=i\omega\mathbf{J}.\label{eq:Lin.Helmholtz.nonop}
\end{equation}
$\mathbf{E}=\mathbf{E}\left(\mathbf{r};\omega\right)$ is the Fourier
component of the electric field at position $\bf r$ and frequency $\omega$, with $c$
the speed of light and $\varepsilon_{0}$ the vacuum permittivity
(we use SI units). On the right hand side of Eq.~\eqref{eq:Lin.Helmholtz.nonop} are the sources \cite{Jackson}, 
which in our case will include the stochastic source for thermal and quantum 
noise, but also a perturbing source to measure the response of the system.
In the following we will consider systems without free charges or currents.

The polarization field $\mathbf{P}$ is a functional of the electric
field $\mathbf{E}$, known as a constituent relation. Since most materials
respond dominantly linearly (unless the electric field is very high),
the polarization is conventionally given in powers of $\mathbf{E}$.
We will consider here materials with a spatially local response, in which case
the polarization at a particular point depends on the electric
field at that same point. A generalization to nonlocal materials is
in principle possible. The polarization vector field is thus
\begin{equation}
\mathbf{P}=\varepsilon_{0}\frac{c^{2}}{\omega^{2}}\left(\mathbb{V}\mathbf{E}+\mathcal{M}\left[\mathbf{E}\otimes\mathbf{E}\right]+\mathcal{N}\left[\mathbf{E}\otimes\mathbf{E}\otimes\mathbf{E}\right]+\dots\right),\label{eq:polarization}
\end{equation}
where the dots represent higher order terms in $\mathbf{E}$. In this
manuscript, we will neglect terms beyond third order. 

Using summation over repeated tensor indices (as used throughout this paper), 
the linear term in Eq.~\eqref{eq:polarization} is given by 
\begin{align}
(\mathbb{V}\mathbf{E})_i=\frac{\omega^{2}}{c^{2}}\chi_{ij}^{\left(1\right)}\left(-\omega,\omega\right)E_{j}\left(\omega\right).\label{eq:V}
\end{align}
In addition to the familiar dielectric function $\varepsilon$, we have introduced
the linear susceptibility $\chi^{\left(1\right)}=\varepsilon-1$ to allow for a consistent
notation of higher orders. For the same reason, we have also kept two frequency 
arguments (for outgoing and incoming waves, which is a standard notation in nonlinear optics), 
the first of which is typically omitted in the linear case because only waves of 
the same frequency interact. The susceptibility depends on a spatial coordinate, 
being zero in vacuum and typically finite and homogeneous inside objects.

The second order term in Eq.~\eqref{eq:polarization} reads
\begin{gather}
\mathcal{M}\left[\mathbf{E}\otimes\mathbf{E}\right]_{i}\left(\omega\right)=\frac{\omega^{2}}{c^{2}}\int d\omega_{1}d\omega_{2}\delta\left(\omega-\omega_{\sigma}\right)\nonumber \\
\times\chi_{ijk}^{\left(2\right)}\left(-\omega_{\sigma},\omega_{1},\omega_{2}\right)E_{j}\left(\omega_{1}\right)E_{k}\left(\omega_{2}\right).\label{eq:M.general}
\end{gather}
The second order susceptibility $\chi^{\left(2\right)}$ carries formally
three frequency arguments. The delta function with $\omega_{\sigma}=\omega_{1}+\omega_{2}$
reflects the fact that the time-domain response depends only on time differences,
i.e., that the susceptibilities are constant in time. The assumption of spatial 
locality implies that it depends on a single spatial coordinate, so that the two 
fields in Eq.~\eqref{eq:M.general} are evaluated at the same position in space. We also introduced the dyadic product '$\otimes$', so that the argument of $\mathcal{M}$ is a second rank spatial tensor.  
The third order term is then a natural extension, 
\begin{align}
\mathcal{N}\left[\mathbf{E}\otimes\mathbf{E}\otimes\mathbf{E}\right]_{i}\left(\omega\right)=\frac{\omega^{2}}{c^{2}}\int d\omega_{1}d\omega_{2}d\omega_{3}\delta\left(\omega-\omega_{\sigma}\right)\nonumber \\
\times\chi_{ijkl}^{\left(3\right)}\left(-\omega_{\sigma},\omega_{1},\omega_{2},\omega_{3}\right)E_{j}\left(\omega_{1}\right)E_{k}\left(\omega_{2}\right)E_{l}\left(\omega_{3}\right).\label{eq:N.general}
\end{align}
Due to intrinsic symmetries in $\chi^{\left(2\right)}$ and 
$\chi^{\left(3\right)}$ \cite{Boyd2008}, the operators
$\mathcal{M}$ and $\mathcal{N}$ are commutative in their operands.
This means $\mathcal{M}\left[\mathbf{A}\otimes\mathbf{B}\right]=\mathcal{M}\left[\mathbf{B}\otimes\mathbf{A}\right]$
and the same applies for permutations in $\mathcal{N}\left[\mathbf{A}\otimes\mathbf{B}\otimes\mathbf{C}\right]$.


Introducing the free Helmholtz operator 
$\mathbb{H}_{0}=\nabla\times\nabla\times-\frac{\omega^{2}}{c^{2}}\mathbb{I}$ and using  
Eq.~\eqref{eq:polarization}, the stochastic nonlinear Helmholtz equation can be written as
\begin{align}
\left(\mathbb{H}_{0}-\mathbb{V}\right)\mathbf{E}-\mathcal{M}\left[\mathbf{E}\otimes\mathbf{E}\right]&-\mathcal{N}\left[\mathbf{E}\otimes\mathbf{E}\otimes\mathbf{E}\right] \nonumber\\
&=\mathbf{F}+\mathbb{H}_{0}\mathbf{E}_{\mathrm{in}},\label{eq:Helmholtz.Full}
\end{align}
where we replaced the generic source term by $i\omega\mathbf{J}=\mathbf{F}+\mathbb{H}_{0}\mathbf{E}_{{\rm in}}$. 
$\mathbf{F}$ is the stochastic source of thermal and quantum noise,
whereas $\mathbf{E}_{{\rm in}}$ denotes a deterministic probing field.
Since we consider a system without free charges or currents, we must have 
$\left\langle\mathbf{F}\right\rangle=0$.  

We mark a few crucial differences between nonlinear and linear Helmholtz equations. First, the 
different frequency components of $\mathbf{E}$ are coupled through Eqs.~\eqref{eq:M.general} 
and \eqref{eq:N.general}. This means fluctuations of all frequencies  
influence the scattering of the electric field of any particular frequency. This is a 
manifestation of the absence of the superposition principle. As a consequence, different frequency components 
cannot be simply added and in general Eq.~\eqref{eq:Helmholtz.Full} 
needs to be solved self-consistently. Our approach is to notice that the nonlinear terms 
are small for most realistic materials and therefore approach the problem perturbatively, 
giving results to leading order in $\chi^{\left(2\right)}$ and $\chi^{\left(3\right)}$.

It is useful to separate the electric field into a
mean part and fluctuations as $\mathbf{E}=\overline{\mathbf{E}}+\delta\mathbf{E}$,
where $\overline{\mathbf{E}}=\mathbf{\left\langle \mathbf{E}\right\rangle }$
is a short hand notation for the average field. In the linear case ($\mathcal{M}=\mathcal{N}=0$)
one obtains two independent equations for $\delta\mathbf{E}$
and $\overline{\mathbf{E}}$, 
\begin{align}
\left(\mathbb{H}_{0}-\mathbb{V}\right)\delta\mathbf{E} & =\mathbf{F},\\
\left(\mathbb{H}_{0}-\mathbb{V}\right)\overline{\mathbf{E}} & =\mathbb{H}_{0}\mathbf{E}_{\mathrm{in}}.
\end{align}
In the next two subsections we will derive equations for the average field
and for the fluctuations in the nonlinear case. We will show that the behavior 
of the average field will depend on the strength of the fluctuations.

\subsection{Average field, effective potential and linear response}

\label{sec:af}  A  linear or nonlinear scattering 
experiment typically detects the noise-averaged field $\overline{\mathbf{E}}$ and its equation of motion shall be derived here.  
As mentioned, due to the absence of the superposition principle, the noise in 
Eq.~\eqref{eq:Helmholtz.Full} has nontrivial consequences for the average field.
This may be seen explicitly by substituting $\mathbf{E}=\overline{\mathbf{E}}+\delta\mathbf{E}$
into Eq.~\eqref{eq:Helmholtz.Full} and taking the average. Using the 
commutative properties of $\mathcal{M}$ and $\mathcal{N}$ together with the fact that by 
definition $\left\langle \delta\mathbf{E}\right\rangle =0$, we obtain a 
nonlinear Helmholtz equation for the mean electric field,
\begin{align}
  &\left(\mathbb{H}_{0}-\mathbb{V}-3\mathcal{N}\left[\left\langle \delta\mathbf{E}\otimes\delta\mathbf{E}\right\rangle \,\cdot\thinspace\right]\right)\overline{\mathbf{E}}\nonumber \\
  &-\mathcal{M}\left[\overline{\mathbf{E}}\otimes\overline{\mathbf{E}}\right]-\mathcal{N}\left[\overline{\mathbf{E}}\otimes\overline{\mathbf{E}}\otimes\overline{\mathbf{E}}\right]\nonumber \\
  &=\mathbb{H}_{0}\mathbf{E}_{\mathrm{in}}+\mathcal{M}\left[\left\langle \delta\mathbf{E}\otimes\delta\mathbf{E}\right\rangle \right]+\mathcal{N}\left[\left\langle \delta\mathbf{E}\otimes\delta\mathbf{E}\otimes\delta\mathbf{E}\right\rangle \right].\label{eq:Helmholtz.average.full}
\end{align}

In stationary systems the time-domain correlators can only depend on time differences \cite{Kubo1966}.
This means that in frequency space the fluctuations are delta-correlated with
$\left\langle \delta\mathbf{E}\left(\omega\right) \otimes\delta\mathbf{E}\left(\omega^{\prime}\right)\right\rangle =\delta\left(\omega+\omega^{\prime}\right)\left\langle\delta\mathbf{E} \otimes\delta\mathbf{E}\right\rangle_{\omega}$. Therefore, the operator $\mathcal{N}\left[\left\langle \delta\mathbf{E}\otimes\delta\mathbf{E}\right\rangle \,\cdot\thinspace\right]$ on the first line of Eq.~\eqref{eq:Helmholtz.average.full} is a linear and local operator (in both position and frequency space) acting on $\overline{\mathbf{E}}$. It can be written explicitly as (we give the spatial argument to emphasize spatial locality)
\begin{gather}
\mathcal{N}\left[\left\langle \delta\mathbf{E}\otimes\delta\mathbf{E} \right\rangle \,\cdot\,\right]_{ij}\left(\mathbf{r};\omega\right)=\notag \\
\label{eq:N.general-1}\frac{\omega^{2}}{c^{2}}\int d\omega'\chi_{ijkl}^{\left(3\right)}\left(\mathbf{r};-\omega,\omega,\omega^{\prime},-\omega^{\prime}\right)\left\langle \delta E_{k}\left(\mathbf{r}\right)\delta E_{l}\left(\mathbf{r}\right)\right\rangle _{\omega^{\prime}}.
\end{gather}
Since it is linear and local, like $\mathbb{V}$ in Eq.~\eqref{eq:V}, we may interpret it as an additional potential, resulting in the effective, fluctuation-dependent potential, 
\begin{equation}
\tilde{\mathbb{V}}=\mathbb{V}+3\mathcal{N}\left[\left\langle \delta\mathbf{E}\otimes\delta\mathbf{E}\right\rangle \,\cdot\,\right].\label{eq:V.general}
\end{equation}
Equivalently, we can define an effective dielectric function corresponding to the
effective potential $\tilde{\mathbb{V}}$ as
\begin{align}
\tilde{\varepsilon}_{ij}\left(\mathbf{r};\omega\right) & =\varepsilon_{ij}\left(\mathbf{r};\omega\right)+\int d\omega^{\prime}N_{ij}\left(\mathbf{r};\omega,\omega^{\prime}\right),\label{eq:epsilon.effective}\\
N_{ij}\left(\mathbf{r};\omega,\omega^{\prime}\right) & =3\chi_{ijkl}^{\left(3\right)}\left(\mathbf{r};-\omega,\omega,\omega^{\prime},-\omega^{\prime}\right)\nonumber \\
 & \times\left\langle \delta E_{k}\left(\mathbf{r}\right)\delta E_{l}\left(\mathbf{r}\right)\right\rangle _{\omega^{\prime}}.\label{eq:N.full}
\end{align}

Moving on to the last line of Eq.~\eqref{eq:Helmholtz.average.full}, we see that in addition to the probing source, two fluctuation-induced sources appear. The source from $\mathcal{M}$ can be written explicitly as  
\begin{gather}
\mathcal{M}\left[\left\langle \delta\mathbf{E}\otimes\delta\mathbf{E}\right\rangle \right]_{i}\left(\mathbf{r};\omega\right)=\delta\left(\omega\right)\frac{\omega^{2}}{c^{2}} \nonumber\\
\times\int d\omega^{\prime}\chi_{ijk}^{\left(2\right)}\left(\mathbf{r};0,\omega^{\prime},-\omega^{\prime}\right)\left\langle \delta E_{j}\left(\mathbf{r}\right)\delta E_{k}\left(\mathbf{r}\right)\right\rangle _{\omega^{\prime}}.\label{eq:M.general-1-1}
\end{gather}
Notably, this term only contributes at $\omega=0$ and may thus 
be interpreted as an average charge generated by fluctuations. It is however 
delicate, as it is in principle in contradiction with the setup of  
$\overline{\mathbf{E}}=0$ as used later. The third order source term, 
$\mathcal{N}\left[\left\langle \delta\mathbf{E}\otimes\delta\mathbf{E}\otimes\delta\mathbf{E}\right\rangle \right]$, 
can not be eliminated generally for any $\omega$. Since the dynamics follows 
a nonlinear equation, the fluctuations $\delta\mathbf{E}$ are generally 
non-Gaussian. It is therefore not obvious how to evaluate the three point correlator. 


Starting from here, we will only keep the \emph{leading order} terms
in  $\chi^{(2)}$ and $\chi^{(3)}$, neglecting terms of order
$\mathcal{O}([\chi^{(2)}]^{2})$,
$\mathcal{O}(\chi^{(2)}\chi^{(3)})$,
$\mathcal{O}([\chi^{(3)}]^{2})$,
and beyond. This simplification is justified from the observation that $\chi^{(3)}$ 
(and $\chi^{(2)}$) are typically small in reality and we aim to calculate the 
leading influence of them on effects like the Casimir force and heat transfer. 
This implies that the fields inside the $\mathcal{M}$ and the $\mathcal{N}$ operator 
in Eq.~\eqref{eq:Helmholtz.average.full} can be found from linear 
theory. Most importantly, the fluctuations $\delta\mathbf{E}$ in this  linear 
system (with $\chi^{(2)}=\chi^{(3)}=0$) are Gaussian. Three point (or any odd number) 
correlations vanish, so that the last term of Eq.~\eqref{eq:Helmholtz.average.full} 
is zero. 
This limit also means that the ``strange'' 
sources appearing in Eq.~\eqref{eq:Helmholtz.average.full} do not couple to finite frequencies and will thus be irrelevant for 
most of the discussion of the paper. Their influence on the Casimir force 
remains however unclear. 

With these considerations, we finally arrive at the nonlinear Helmholtz
equation for the average field $\overline{\mathbf{E}}$ (valid for $\omega\not=0$),
\begin{equation}
\left(\mathbb{H}_{0}-\tilde{\mathbb{V}}\right)\overline{\mathbf{E}}
-\mathcal{M}\left[\overline{\mathbf{E}}\otimes\overline{\mathbf{E}}\right]-
\mathcal{N}\left[\overline{\mathbf{E}}\otimes\overline{\mathbf{E}}\otimes\overline{\mathbf{E}}\right]=\mathbb{H}_{0}\mathbf{E}_{\mathrm{in}}.\label{eq:Helmholtz.average.effective}
\end{equation}
We recover the same structure as in Eq.~\eqref{eq:Helmholtz.Full}, only without the noise
term (making it deterministic) and a modified, renormalized, linear term. 
This equation determines the 
result of both linear and nonlinear scattering experiments. Furthermore, we note 
that $\chi^{(2)}$ and 
$\chi^{(3)}$ (entering $\mathcal{M}$ and 
$\mathcal{N}$) are 
not renormalized by the noise in this order. That would appear by 
inclusion of $\chi^{(5)}$ and so on.

To make even clearer the meaning of the effective potential (and the effective
dielectric function), we explicitly compute the result of a linear
response scattering experiment. To that end we interpret $\mathbf{E}_{\mathrm{in}}$
as the incident field in such an experiment and find the scattered one
linear in it, thereby defining the linear response function $\tilde{\mathbb{G}}$,
\begin{equation}
\tilde{\mathbb{G}}=\left.\frac{\delta\overline{\mathbf{E}}}{\delta\left(\mathbb{H}_{0}\mathbf{E}_{\mathrm{in}}\right)}\right|_{\mathbf{E}_{\mathrm{in}}\rightarrow0}=\left.\frac{\delta\overline{\mathbf{E}}}{\delta\mathbf{E}_{\mathrm{in}}}\right|_{\mathbf{E}_{\mathrm{in}}\rightarrow0}\mathbb{G}_{0}\label{eq:G.definition},
\end{equation}
with the free Green's function $\mathbb{G}_{0}=\mathbb{H}_{0}^{-1}$
\cite{Jackson}. $\tilde{\mathbb{G}}$ follows directly from Eq.~\eqref{eq:Helmholtz.average.effective}
as  \cite{Soo2016a}
\begin{equation}
\tilde{\mathbb{G}}=\left(\mathbb{H}_{0}-\tilde{\mathbb{V}}\right)^{-1}.\label{eq:G.general}
\end{equation}
The linear response is indeed given by the effective potential $\tilde{\mathbb{V}}$
or, equivalently, by the effective dielectric function 
$\tilde{\varepsilon}$, which depends on the fluctuations through Eq.~\eqref{eq:N.full}. 

Notably, for $\chi^{(3)}=0$, we recover the well known linear response
function. In that case Eq.~\eqref{eq:G.general} reduces to the
Green's function $\mathbb{G}=\left(\mathbb{H}_{0}-\mathbb{V}\right)^{-1}$
of the system \cite{rahi2009scattering}. We will see in the following
that it is the effective potential, which will give rise to effects
of $\chi^{(3)}$ on Casimir forces and heat transfer. Analogously to Eq.~\eqref{eq:G.definition},
higher order derivatives can be used to infer the nonlinear susceptibilities,
but they will not be relevant for the remainder of the paper, because they play no role in fluctuational effects.

\subsection{Equilibrium fluctuations and Rytov currents\label{subsec:Equilibrium-fluctuations-and}}

In the previous subsections we have determined Eq.~(\ref{eq:Helmholtz.Full}) for the
fluctuating field $\mathbf{E}$ and Eq.~(\ref{eq:Helmholtz.average.effective})
for its mean $\overline{\mathbf{E}}$. However,  in the first case we need to know
the noise source $\mathbf{F}$ and in the second case the correlations of the 
fluctuations $\delta\mathbf{E}$ in Eq.~\eqref{eq:V.general}. Because these equations are
linked by $\mathbf{E}=\overline{\mathbf{E}}+\delta\mathbf{E}$ and
therefore $\left\langle \mathbf{E}\otimes\mathbf{E}\right\rangle =\overline{\mathbf{E}}\otimes\overline{\mathbf{E}}+\left\langle \delta\mathbf{E}\otimes\delta\mathbf{E}\right\rangle $,
we will now determine  $\delta\mathbf{E}$ by use of the FDT.
We will then find the correlations of $\mathbf{F}$ and connect them to Rytov theory \cite{Levin1967,Rytov3}.


The equilibrium fluctuations $\left\langle \delta\mathbf{E} \otimes\delta\mathbf{E}\right\rangle^{\mathrm{eq}}$ are related 
to the linear response function $\tilde{\mathbb{G}}$ defined by Eq.~\eqref{eq:G.definition}
through the FDT \cite{Kubo1966}, given explicitly as 
\begin{align}
\left\langle \delta\mathbf{E}_{\omega}\otimes\delta\mathbf{E}_{\omega^{\prime}}^{*}\right\rangle ^{\mathrm{eq}}&=\delta\left(\omega-\omega^{\prime}\right)b\left(\omega\right)\mathrm{Im}\tilde{\mathbb{G}}^{\mathrm{eq}}\left(\omega\right),\label{eq:FDT}\\
b\left(\omega\right)&=\frac{\hbar}{\pi\varepsilon_{0}}\frac{\omega^{2}}{c^{2}}\frac{1}{1-\exp\left(-\frac{\hbar\omega}{k_{\mathrm{B}}T}\right)},
\end{align}
with the reduced Planck constant $\hbar$ and thermal energy $k_{\mathrm{B}}T$.
$\tilde{\mathbb{G}}^{\mathrm{eq}}=\left(\mathbb{H}_{0}-\tilde{\mathbb{V}}^{\mathrm{eq}}\right)^{-1}$
is the linear response, given by Eq.~\eqref{eq:G.general}, in equilibrium, which is denoted by the superscript 'eq', also used for averages $\langle\dots\rangle^{\rm eq}$. Unlike in linear systems, the linear response $\tilde{\mathbb{G}}$
as well as the effective potential $\tilde{\mathbb{V}}$ in Eq.~\eqref{eq:V.general}
depend on the correlations of $\delta\mathbf{E}$ and are in general different
in equilibrium as compared to the  nonequilibrium situation  considered in Sec.~\ref{sec:Thermal-inequilibrium}. 

The effective potential or dielectric function in equilibrium is given by Eqs.~\eqref{eq:V.general}
or \eqref{eq:N.full} with the equilibrium correlator on the rhs. This will be discussed in Subsection~\ref{subsec:The-effective-dielectric} below. Eq.~\eqref{eq:FDT}
gives a rigid and well-known relation between two different measurable
quantities, and is the heart of our analysis. The correlator $\left\langle \delta\mathbf{E}\otimes\delta\mathbf{E}\right\rangle ^{\mathrm{eq}}$
determines the Casimir force, while the linear response $\tilde{\mathbb{G}}^{\mathrm{eq}}$ describes optical  scattering experiments.

Using the fluctuations obtained by the FDT, we can also determine the correlations
of the noise sources $\mathbf{F}$. For $\overline{\mathbf{E}}=\mathbf{E}_{\mathrm{in}}=0$,
we have from Eq.~\eqref{eq:Helmholtz.Full} (we include here a finite $\chi^{(2)}$ to demonstrate its consequences)
\begin{equation}
\mathbf{F}=\left(\mathbb{H}_{0}-\mathbb{V}\right)\delta\mathbf{E}-\mathcal{M}\left[\delta\mathbf{E}\otimes\delta\mathbf{E}\right]-\mathcal{N}\left[\delta\mathbf{E}\otimes\delta\mathbf{E}\otimes\delta\mathbf{E}\right].\label{eq:F}
\end{equation}
As before, we note a subtlety at zero frequency for finite $\chi^{\left(2\right)}\left(0,\omega^\prime,-\omega^\prime\right)$. This implies a contradiction of Eq.~\eqref{eq:F} with the fundamental assumption $\langle\mathbf{F}\rangle=0$, however, only at $\omega=0$. 

Keeping only terms up to leading order in $\chi^{\left(2\right)}$ and 
$\chi^{\left(3\right)}$, the equilibrium correlator of $\mathbf{F}$ follows directly from Eq.~\eqref{eq:F}, 
\begin{align}
\left\langle \mathbf{F}_{\omega}\otimes\mathbf{F}_{\omega^{\prime}}^{*}\right\rangle ^{\mathrm{eq}}&=\left(\mathbb{H}_{0}-\mathbb{V}\right)_\omega\left\langle \delta\mathbf{E}_{\omega}\otimes\delta\mathbf{E}_{\omega^{\prime}}^{*}\right\rangle ^{\mathrm{eq}}\left(\mathbb{H}_{0}-\mathbb{V}\right)^{*}_{\omega^\prime}\nonumber\\
&\hspace{-4em}-\left(\mathbb{H}_{0}-\mathbb{V}\right)_{\omega}\left\langle \delta\mathbf{E}_{\omega}\otimes\mathcal{N}\left[\delta\mathbf{E}\otimes\delta\mathbf{E}\otimes\delta\mathbf{E}\right]_{\omega^{\prime}}^{*}\right\rangle ^{\mathrm{eq}}\nonumber\\
&\hspace{-4em}-\left\langle \mathcal{N}\left[\delta\mathbf{E}\otimes\delta\mathbf{E}\otimes\delta\mathbf{E}\right]_{\omega}\otimes\delta\mathbf{E}_{\omega^{\prime}}^{*}\right\rangle ^{\mathrm{eq}}\left(\mathbb{H}_{0}-\mathbb{V}\right)^{*}_{\omega^\prime}.
\end{align}
We can further assume Gaussianity of the fields in the last two terms, because they carry already an explicit factor of $\chi^{(3)}$ (and are thus to be taken from the linear system). Specifically, by using Isserlis' theorem
and the commutation properties of $\mathcal{N}$, we can write
\begin{align}
\left\langle \mathcal{N}\left[\delta\mathbf{E}\otimes\delta\mathbf{E}\otimes\delta\mathbf{E}\right]\otimes\delta\mathbf{E}^{*}\right\rangle^\mathrm{eq} &=3\mathcal{N}\left[\left\langle \delta\mathbf{E}\otimes\delta\mathbf{E}\right\rangle^\mathrm{eq} \,\cdot\thinspace\right] \nonumber\\
&\hspace{-0em}\times\left\langle \delta\mathbf{E}\otimes\delta\mathbf{E}^{*}\right\rangle^\mathrm{eq} .
\end{align}
We note the appearance of the same operator as in Eq.~\eqref{eq:V.general},
which may be written in terms of $\tilde{\mathbb{V}}$ or, equivalently,
in terms of $\tilde{\mathbb{G}}$. We therefore find 
\begin{align}
\left\langle \mathbf{F}_{\omega}\otimes\mathbf{F}_{\omega^{\prime}}^{*}\right\rangle ^{\mathrm{eq}} & =\left(\tilde{\mathbb{G}}^{\mathrm{eq}}\right)_{\omega}^{-1}\left\langle \delta\mathbf{E}_{\omega}\otimes\delta\mathbf{E}_{\omega^{\prime}}^{*}\right\rangle ^{\mathrm{eq}}\left(\tilde{\mathbb{G}}^{\mathrm{eq}*}\right)_{\omega^{\prime}}^{-1}.\label{eq:FF.ee}
\end{align}
Using Eqs. \eqref{eq:G.general} and \eqref{eq:FDT}, we can further write
this as
\begin{align}
\left\langle \mathbf{F}_{\omega}\otimes\mathbf{F}_{\omega^{\prime}}^{*}\right\rangle ^{\mathrm{eq}} & =-\delta\left(\omega-\omega^{\prime}\right)b\left(\omega\right)\mathrm{Im}\left(\mathbb{H}_{0}-\tilde{\mathbb{V}}^{\mathrm{eq}}\right)_{\omega},\label{eq:Rytov.currents.equilibrium}
\end{align}
which matches with the relation for Rytov currents for linear systems
\cite{Rytov3,Levin1967}. As noted in Ref.~\cite{Soo2016a}, the
noise sources (the Rytov currents) are related to the effective potential
$\tilde{\mathbb{V}}^{\mathrm{eq}}$ in the same manner as they are
related to the bare potential $\mathbb{V}$ in linear systems. This confirms the interpretation of  $\tilde{\mathbb{V}}^{\mathrm{eq}}$ as the linear response function, as it appears in a fluctuation dissipation theorem with the noise in Eq.~\eqref{eq:Rytov.currents.equilibrium}. Eqs.~\eqref{eq:FDT} and \eqref{eq:Rytov.currents.equilibrium} are thus two versions of the fluctuation dissipation theorem \cite{Kubo1966}.


We have thus demonstrated the consistency of FDT and Rytov theory in
the nonlinear case. It is interesting to note that the Rytov
currents are uncorrelated in space, as they are in linear systems,
because the effective potential is local in space. The potential and
the noise are however nonlocal in the sense that their value at one
position depends on the properties of the system at all other points
in space. For example, the effective potential of a point inside an
object depends on the shape of the object or on the presence of surrounding
objects. It means that Eq. \eqref{eq:Rytov.currents.equilibrium}
is an implicit equation for $\left\langle \mathbf{F}\otimes\mathbf{F}^{*}\right\rangle ^{\mathrm{eq}}$, 
just as Eq.~\eqref{eq:FDT} is an implicit equation for $\left\langle \delta\mathbf{E}\otimes\delta\mathbf{E}^{*}\right\rangle ^{\mathrm{eq}}$.

We note that $\mathrm{Im}\tilde{\mathbb{V}}^{\mathrm{eq}}$ in Eq.~\eqref{eq:Rytov.currents.equilibrium} must be positive, as can already be seen by its connection to an autocorrelation function. This property is however hard to show explicitly without posing additional constraints on $\chi^{(3)}$.

\subsection{The effective dielectric function in equilibrium\label{subsec:The-effective-dielectric}}

In previous sections we showed how the effective potential $\tilde{\mathbb{V}}$ 
can be used to take into account nonlinear effects in equilibrium. The most 
important quantity is the linear response function, equivalently expressed by 
the potential  $\tilde{\mathbb{V}}^{\mathrm{eq}}$, the dielectric function 
$\tilde{\varepsilon}_{ij}^{\mathrm{eq}}$, or $\tilde{\mathbb{G}}^{\rm eq}$, 
because it governs the fluctuations. We will thus investigate the linear response
in more detail with simple examples in this section.

Writing Eqs.~\eqref{eq:epsilon.effective} and \eqref{eq:N.full}
using the FDT in Eq. \eqref{eq:FDT} gives the effective
dielectric function in equilibrium
\begin{align}
\tilde{\varepsilon}_{ij}^{\mathrm{eq}}\left(\mathbf{r};\omega\right) & =\varepsilon_{ij}\left(\mathbf{r};\omega\right)+\int d\omega^{\prime}N_{ij}^{\mathrm{eq}}\left(\mathbf{r};\omega,\omega^{\prime}\right), \label{eq:epsilon.effective-eq} \\
N_{ij}^{\mathrm{eq}}\left(\mathbf{r};\omega,\omega^{\prime}\right) & =3\chi_{ijkl}^{\left(3\right)}\left(\mathbf{r};-\omega,\omega,\omega^{\prime},-\omega^{\prime}\right)\nonumber \\
 & \times b\left(\omega^{\prime}\right)\mathrm{Im}\mathbb{G}\left(\mathbf{r},\mathbf{r};\omega^{\prime}\right)_{kl}^{\mathrm{eq}}. \label{eq:N.eq.full}
\end{align}
Note that we used the Green's function $\mathbb{G}=\left(\mathbb{H}_{0}-\mathbb{V}\right)^{-1}$ instead of the linear response (with a tilde) as in Eq.~\eqref{eq:FDT}. 
This is correct to leading order in $\chi^{(3)}$. $\mathbb{G}$ is known exactly for
several geometries, so that these equations are closed. 

We note that the imaginary part of the Green's function, as appearing
in Eq.~\eqref{eq:N.eq.full}, generally diverges at coinciding points
in absorbing media, which is a well known property \cite{Jackson} and a recurrent problem of perturbative
expansions in field theories \cite{kardarbook}. There have been suggestions
on how to circumvent this divergence, e.g. by introducing a rigid sphere approximation of the delta function \cite{Matloob1999} appearing in the Green's function, which appears very similar
to an ultraviolet cut-off often introduced in classical
field theory.

This problem can also be mitigated in some cases.
For example, when computing the Casimir force in Ref.~\cite{Soo2016a},
we noted that the nontrivial distance dependence of the force is not 
sensitive to the divergence, as it cancels out when comparing two different 
distances. It is thus important to carefully investigate  which experimental 
quantities are insensitive to the mentioned divergence and can thus be predicted.
In the remainder of the paper we will point to this issue for any
shown example and reflect on it in the summary section. We also comment
that using a purely real $\varepsilon$ omits the divergence in any circumstance.

In the interest of simplifying the calculation and interpretation of specific 
examples,  especially in view of 
the complicated tensorial structure of $\chi_{ijkl}^{\left(3\right)}$, we consider a highly symmetric material. First, 
we assume that the bare dielectric function is isotropic, such
that $\varepsilon_{ij}=\delta_{ij}\varepsilon$. Regarding $\chi_{ijkl}^{\left(3\right)}$,
it is known that for centro-symmetric materials the third order susceptibility
can be written as \cite{Boyd2008} 
\begin{equation}
\chi_{ijkl}^{\left(3\right)}=\chi_{1122}^{\left(3\right)}\delta_{ij}\delta_{kl}+\chi_{1212}^{\left(3\right)}\delta_{ik}\delta_{jl}+\chi_{1221}^{\left(3\right)}\delta_{il}\delta_{jk}.\label{eq:sym}
\end{equation}
Further simplifying,
we only keep the first term from Eq. \eqref{eq:sym}, such that we use
\begin{align}
\chi_{ijkl}^{\left(3\right)}=\chi^{\left(3\right)}\delta_{ij}\delta_{kl}.
\end{align}
With these simplifications, the resulting effective dielectric function is isotropic,
\begin{align}
\tilde{\varepsilon}_{ij}^{\mathrm{eq}}\left(\mathbf{r};\omega\right)&=\delta_{ij}\left[\varepsilon\left(\mathbf{r};\omega\right)+\int d\omega^{\prime}N^{\mathrm{eq}}\left(\mathbf{r};\omega,\omega^{\prime}\right)\right],\label{eq:epsilon.eq.iso} \\
N^{\mathrm{eq}}\left(\mathbf{r};\omega,\omega^{\prime}\right)
&=3\chi^{\left(3\right)}\left(\mathbf{r};-\omega,\omega,\omega^{\prime},-\omega^{\prime}\right) \nonumber \\
&\times b\left(\omega^{\prime}\right)\mathrm{Im}\mathbb{G}\left(\mathbf{r},\mathbf{r};\omega^{\prime}\right)_{kk}^{\mathrm{eq}}. \label{eq:N.eq.iso}
\end{align}

We will consider the examples of a single plate and two parallel plates using 
Ref.~\cite{Johansson2011}. It gives $\mathbb{G}$ in plane wave basis for 
arbitrary parallel layered structures, which contains the cases of a single 
semi-infinite plate (two layers: vacuum--plate) and two parallel semi-infinite 
plates (three layers: plate--vacuum--plate).

\begin{figure}
\includegraphics[width=1\columnwidth]{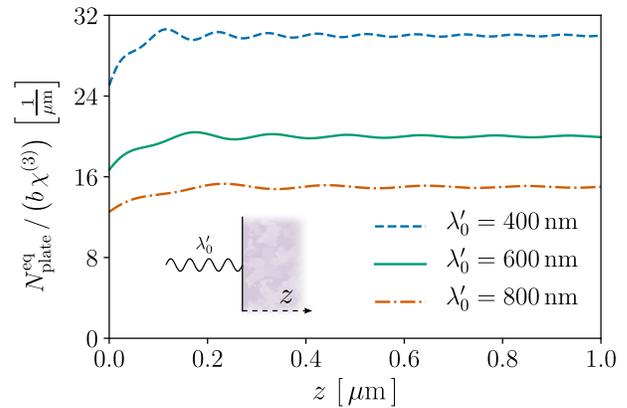} \caption{Spatial dependence of the effective 
dielectric function inside a single isotropic non-absorbing plate ($\varepsilon=4$). 
$z$ is the distance from the surface and $\lambda_{0}^{\prime}=2\pi\frac{c}{\omega^{\prime}}$
is the wavelength corresponding to $\omega^{\prime}$ in vacuum. Note that while
$N_{\mathrm{plate}}^{\mathrm{eq}}\left(\mathbf{r};\omega,\omega^{\prime}\right)$
depends also on $\omega$, the plotted quantity is independent of $\omega$ due to 
division with $\chi^{\left(3\right)}\left(\mathbf{r};-\omega,\omega,\omega^{\prime},-\omega^{\prime}\right)$.\label{fig:N.single.eq}}
\end{figure}
We start with a single plate. As mentioned above, the imaginary part 
of the Green's function at coinciding points [$\mathrm{Im}\tilde{\mathbb{G}}\left(\mathbf{r},\mathbf{r};\omega^{\prime}\right)$, as appearing in Eq.~\eqref{eq:N.eq.iso}] is infinite inside absorbing materials, so the effective dielectric function cannot 
be computed without further (microscopic) information in general. Subtracting from 
it the solution of an unbound (bulk) system with the same dielectric
function heals the divergence, except for points which are very close
to the plate's surface. In the case of a single semi-infinite plate, 
we therefore restrict ourselves to a real bare $\varepsilon=4$, for which Eq.~\eqref{eq:N.eq.iso} can be numerically evaluated. The result, which is nevertheless insightful,
is shown in Fig.~\ref{fig:N.single.eq}. The effective dielectric
function is inhomogeneous even though all material parameters ($\chi^{(1)}$
and $\chi^{(3)}$) are homogeneous. We thus see explicitly the aforementioned
property: The effective dielectric function at a certain position
depends on the shape of the object. Specifically, there is an interference pattern 
of half of the wavelength of the primed frequency, which corresponds to a single reflection 
from the surface, while far away from the surface the bulk value is approached. Note, 
however, that since in Eq.~\eqref{eq:epsilon.eq.iso} we integrate over all $\omega^\prime$, 
this interference pattern appears in the effective dielectric function only if  
$\chi^{\left(3\right)}\left(-\omega,\omega,\omega^{\prime},-\omega^{\prime}\right)$ 
has a sharp resonance peak in $\omega^\prime$.


The case of two identical parallel surfaces at distance $d$ provides additional insights.  
Here, we may compute the difference of the effective dielectric function between the cases of the second plate present or absent. Mathematically, we thus subtract from $N^\mathrm{eq}_\mathrm{double}$ 
the result of the isolated plate, $N^\mathrm{eq}_\mathrm{plate}$, and obtain a finite result, because the divergence in $\mathbb{G}$ cancels. This important 
observation, which was also used in Ref.~\cite{Soo2016a} to compute the Casimir force, contains a very important physical statement: While it is difficult to predict the response of a single object, it is possible to predict the response of two objects, given the 
responses of the individual objects are known.

\begin{figure}
\includegraphics[width=1\columnwidth]{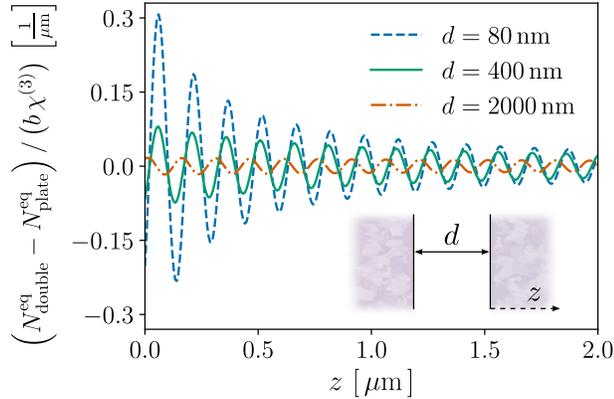} 
\caption{The effective dielectric function inside one of two non-absorbing parallel 
plates ($\varepsilon=4$), where $z$ is the distance from the surface and $d$ is the 
separation between the plates. $\lambda_{0}^{\prime}=2\pi\frac{c}{\omega^{\prime}}=600\,\mathrm{nm}$ 
is the corresponding wavelength in vacuum. \label{fig:N.double.eq}}
\end{figure}

\begin{figure}
\includegraphics[width=1\columnwidth]{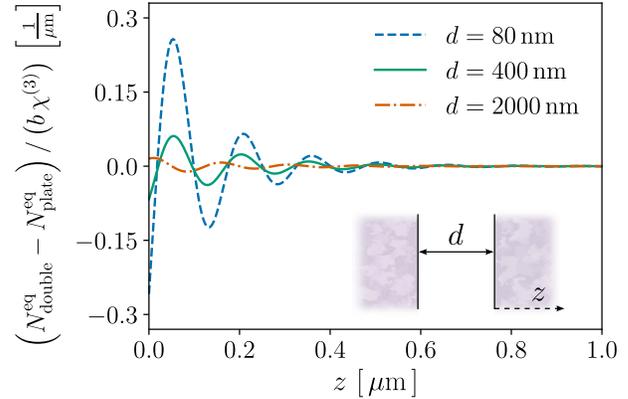} 
\caption{Same quantity as in Figure \ref{fig:N.double.eq} for lossy plates
with a dielectric constant $\varepsilon=4+i$ and vacuum wavelength 
$\lambda_{0}^{\prime}=2\pi\frac{c}{\omega^{\prime}}=600\,\mathrm{nm}$.
\label{fig:N.double.eq.absorbing}}
\end{figure}

The numerical results for identical plates are shown in Figures \ref{fig:N.double.eq}
(without absorption) and \ref{fig:N.double.eq.absorbing} (with absorption).
Not unlike in the single plate case, an interference pattern arises due to
reflections from the second plate (notice the phase shift at different
separations). In the non-absorbing case these persist throughout the
material, while they are limited by the skin depth in absorbing materials.
As expected, at large separations $d$ we recover the single plate result.

These graphs have a very direct connection to the Casimir force between
two parallel nonlinear plates of Ref.~\cite{Soo2016a}. As mentioned
before, the Casimir force is now found by using the well known Lifshitz
formula \cite{Lifshitz56} for linear materials, but replacing the
bare $\varepsilon$ by the (inhomogeneous) effective one. Recall that in Ref.~\cite{Soo2016a},
we found that the force displays a different power law as a function
of $d$ for close separations. For the quantum limit the power law
changes as $d^{-4}\rightarrow d^{-8}$ and for the thermal limit $d^{-3}\rightarrow d^{-6}$.
This may now be understood in terms of Figs.~\ref{fig:N.double.eq}
and \ref{fig:N.double.eq.absorbing}, because the effective dielectric function 
changes with $d$, yielding an additional $d$ dependence in the Casimir force.

We stress again that these statements regarding the force do not take into account the ``strange'' charge of Eq.~\eqref{eq:M.general-1-1}, so that they are strictly true if $\chi^{\left(2\right)}\left(0,\omega^\prime,-\omega^\prime\right)=0$. The force for a finite $\chi^{\left(2\right)}\left(0,\omega^\prime,-\omega^\prime\right)$ needs to be investigated in the future.

\section{Nonequilibrium: Heat radiation\label{sec:Thermal-inequilibrium}}

\subsection{Nonequilibrium Rytov currents and field correlations}\label{sec:NeqR}

In Section \ref{sec:Nonlinear-luctuational-electrody} we developed
FE for nonlinear materials in equilibrium, from which
inhomogeneous dielectric functions and the Casimir effect for nonlinear
objects can be found. In equilibrium, the theory is well grounded
by the FDT, relating the linear response and field fluctuations directly
by Eq. \eqref{eq:FDT}. We now aim to address the out of equilibrium
scenario of $N$ objects held at different temperatures $T_{n}$
($n=1\dots N$), with an environment at temperature $T_{0}$, and compute
heat radiation and heat transfer for these objects.

In this nonequilibrium case, certain assumptions are necessary to compute
the fluctuations of the electric field, because the FDT is in general not 
valid. A useful approximation, which is also used for linear FE, is
the assumption of local thermal equilibrium (LTE). In this case, the (non-overlapping) objects are considered to be in thermal equilibrium at temperatures $T_n$. 


To be able to assign different temperatures,  we start by denoting the susceptibilities of order $m$ ($m$=1,  3) 
of object $n$ as $\chi_{n}^{(m)}({\bf r})$. These are nonzero only when
$\mathbf{r}$ lies within object $n$. Note that $\chi^{(2)}$ has no influence on the following discussion.
Since the objects are spatially separated from 
one another, the total susceptibilities can be found through summation as
\begin{equation}
\chi^{\left(m\right)}\left(\mathbf{r}\right)=\sum_{n=1}^{N}\chi_{n}^{\left(m\right)}\left(\mathbf{r}\right).
\end{equation}
We may also write $\chi^{\left(m\right)}\left(\mathbf{r}\in V_{n}\right)=\chi_{n}^{\left(m\right)}\left(\mathbf{r}\right)$,
where $V_{n}$ is the volume of the object. The same applies for the bare
potential ($\mathbb{V}=\sum_{n=1}^{N}\mathbb{V}_{n}$) and the nonlinear
operator ($\mathbb{\mathcal{N}}=\sum_{n=1}^{N}\mathcal{N}_{n}$), as they follow from $\chi^{\left(1\right)}$ and $\chi^{\left(3\right)}$, respectively. The effective 
potential is then $\tilde{\mathbb{V}}=\sum_{n=1}^{N}\tilde{\mathbb{V}}_{n}$, where we 
have [see Eqs.~\eqref{eq:N.general-1} and \eqref{eq:V.general}]
\begin{equation}
\tilde{\mathbb{V}}_{n}=\mathbb{V}_{n}+3\mathcal{N}_{n}\left[\left\langle \delta\mathbf{E}\otimes\delta\mathbf{E}\right\rangle \,\cdot\thinspace\right].\label{eq:Vn.general}
\end{equation}

The key point in implementing the LTE approximation within FE is to recognize 
that the equilibrium Rytov currents in Eq.~\eqref{eq:Rytov.currents.equilibrium} can be written as 
\begin{align}
\left\langle \mathbf{F}\otimes\mathbf{F}^{*}\right\rangle _{\omega}^{\mathrm{eq}} & =-\mathrm{Im}\left[b\left(\omega\right)\mathbb{H}_{0}-\sum_{n=1}^{N}b\left(\omega\right)\tilde{\mathbb{V}}_{n}^{\mathrm{eq}}\right].\label{eq:Rytov.currents.equilibrium-1}
\end{align}
Since object $n$ is considered to be in local equilibrium at temperature $T_{n}$,
we can assign an index to the distributions $b\left(\omega\right)$, 
\begin{equation}
b_{n}\left(\omega\right)=\frac{\hbar}{\pi\varepsilon_{0}}\frac{\omega^{2}}{c^{2}}\left[1-\exp\left(-\frac{\hbar\omega}{k_{\mathrm{B}}T_{n}}\right)\right]^{-1}.
\end{equation}
Recalling that index $n=0$ denotes the environment with temperature
$T_{0}$, we arrive at the nonequilibrium correlator of the Rytov currents,
\begin{equation}
\left\langle \mathbf{F}\otimes\mathbf{F}^{*}\right\rangle _{\omega}=\sum_{n=0}^{N}b_{n}\left(\omega\right)\mathrm{Im}\left[\tilde{\mathbb{V}}_{n}\right].\label{eq:Fn}
\end{equation}
For brevity, we have denoted $\mathbb{V}_{0}=\tilde{\mathbb{V}}_{0}=-\mathbb{H}_{0}$
as the vacuum potential, which is usually regarded as the environment
dust \cite{Eckhardt1984}. Note that the effective potential $\tilde{\mathbb{V}}$
depends on the field fluctuations $\left\langle \delta\mathbf{E}\otimes\delta\mathbf{E}\right\rangle $
and is thus different out of equilibrium compared to the
corresponding equilibrium potential $\tilde{\mathbb{V}}^{\mathrm{eq}}$.

The correlations of the field fluctuations $\delta\mathbf{E}$ out
of equilibrium can be found by following the same reasoning leading
to Eq. \eqref{eq:FF.ee} (with $\overline{\mathbf{E}}=0$), giving
\begin{align}
&\left\langle \delta\mathbf{E}_{\omega}\otimes\delta\mathbf{E}_{\omega^{\prime}}^{*}\right\rangle =\tilde{\mathbb{G}}_{\omega}\left\langle \mathbf{F}_{\omega}\otimes\mathbf{F}_{\omega^{\prime}}^{*}\right\rangle \tilde{\mathbb{G}}_{\omega^{\prime}}^{*},\nonumber \\
&=\delta\left(\omega-\omega^{\prime}\right)\sum_{n=0}^{N}b_{n}\left(\omega\right)\left(\tilde{\mathbb{G}}\mathrm{Im}\left[\tilde{\mathbb{V}}_{n}\right]\tilde{\mathbb{G}}^{*}\right)_{\omega}.\label{eq:ee.FF}
\end{align}
Recall that the linear response operator is given by Eq.~\eqref{eq:G.general}
as $\tilde{\mathbb{G}}=(\mathbb{H}_{0}-\tilde{\mathbb{V}})^{-1}$ together with Eq.~\eqref{eq:Vn.general}, so that Eq.~\eqref{eq:Fn} is indeed physically meaningful, because the potential $\tilde{\mathbb{V}}$ is the physical linear response of the nonequilibrium system.  

As in the equilibrium case, we have an implicit system of equations to determine 
the fluctuations and the effective potential. It can be solved perturbatively 
in $\chi^{(3)}$ and we will derive an explicit form for the effective 
dielectric function in the next subsection, from which the correlator 
$\left\langle \delta\mathbf{E}\otimes\delta\mathbf{E}^{*}\right\rangle$ 
can then be computed with Eq.~\eqref{eq:ee.FF}.

\subsection{The nonequilibrium effective dielectric function}

From Eqs.~\eqref{eq:Fn} and \eqref{eq:ee.FF} we can see that, as in the equilibrium 
case, the effects of nonlinear terms on fluctuations can be taken into
account with the effective potential $\tilde{\mathbb{V}}$ or, equivalently, the effective
dielectric function. The latter is obtained by substituting the correlator 
for the electric field fluctuations of Eq.~\eqref{eq:ee.FF} into Eqs.~\eqref{eq:epsilon.effective}
and \eqref{eq:N.full}, giving us
\begin{align}
\tilde{\varepsilon}_{ij}\left(\mathbf{r};\omega\right) & =\varepsilon_{ij}\left(\mathbf{r};\omega\right)+\int d\omega^{\prime}N_{ij}\left(\mathbf{r};\omega,\omega^{\prime}\right),\label{eq:epsilon.effective-noneq}\\
N_{ij}\left(\mathbf{r};\omega,\omega^{\prime}\right) & =\sum_{m=0}^{N}3\chi_{ijkl}^{\left(3\right)}\left(\mathbf{r};-\omega,\omega,\omega^{\prime},-\omega^{\prime}\right)\nonumber \\
 & \hspace{-2em} \times b_{m}\left(\omega^{\prime}\right)\left(\tilde{\mathbb{G}}\mathrm{Im}\left[\tilde{\mathbb{V}}_{m}\right]\tilde{\mathbb{G}}^{*}\right)\left(\mathbf{r},\mathbf{r};\omega^{\prime}\right)_{kl}.\label{eq:N.full-noneq}
\end{align}
These expressions reduce to the equilibrium cases $\tilde{\varepsilon}_{ij}^{\mathrm{eq}}$
and $N_{ij}^{\mathrm{eq}}$ of Eqs. \eqref{eq:epsilon.effective-eq}
and \eqref{eq:N.eq.full} if all temperatures are equal. This is because, by definition, 
\begin{equation}
\sum_{m=0}^{\infty}\tilde{\mathbb{V}}_{m}=-\tilde{\mathbb{G}}^{-1}.
\end{equation}
Note that the sum starts at 0, so that it contains also the famous environment dust \cite{Eckhardt1984}. If all temperatures are equal then one recovers $\sum_{m}b\left(\omega^{\prime}\right)\tilde{\mathbb{G}}\mathrm{Im}\left[\tilde{\mathbb{V}}_{m}\right]\tilde{\mathbb{G}}^{*}=b\left(\omega^{\prime}\right)\mathrm{Im}\tilde{\mathbb{G}}$, as in the equilibrium expression.

It is instructive and useful to isolate the nonequilibrium contribution of the
effective dielectric function. It is defined as 
\begin{align}
N^{\mathrm{neq}} & =N-N^{\mathrm{eq}},
\end{align}
where $N^{\mathrm{eq}}$ is the equilibrium limit corresponding to the temperature at the position where $N$ is evaluated.
For ${\bf r}$ located inside object $n$, it reads 
\begin{align}
N_{ij}^{\mathrm{neq}}\left(\mathbf{r}\in V_{n};\omega,\omega^{\prime}\right) & =3\chi_{ijkl}^{\left(3\right)}\left(\mathbf{r};-\omega,\omega,\omega^{\prime},-\omega^{\prime}\right)\label{eq:N.neq.full}\\
 & \times\sum_{m=0}^{N}\left[b_{m}\left(\omega^{\prime}\right)-b_{n}\left(\omega^{\prime}\right)\right]\nonumber \\
 & \times\left(\tilde{\mathbb{G}}\mathrm{Im}\left[\tilde{\mathbb{V}}_{m}\right]\tilde{\mathbb{G}}^{*}\right)\left(\mathbf{r},\mathbf{r};\omega^{\prime}\right)_{kl}.\nonumber 
\end{align}
This expression depends on the temperatures of all objects, because the nonlinear term couples the fluctuations in the different objects.

More precisely, in the above expression only objects with $T_{m}\not=T_{n}$
contribute, where $T_{n}$ is the temperature at $\mathbf{r}$. 
This has an important implication regarding the mentioned divergence of $\mathbb{G}$ at coinciding points. Because $\tilde{\mathbb{V}}_{m}$ is only non-zero
inside body $m$ and ${\bf r}$ is
inside body $n$, the two Green's functions in Eq.~\eqref{eq:N.neq.full}
connect points in \textit{different} objects only (the sum does not
contain the term $m=n$). The expression for ${\mathbb{G}}$
evaluated at two different points is notably finite. We thus find
that the deviation of the effective dielectric function from its equilibrium
value is a quantity which can be  predicted within this framework.

If we have only a single body in vacuum, then the above expression simplifies to
\begin{align}
N_{ij,\mathrm{single}}^{\text{neq}}\left(\mathbf{r};\omega,\omega^{\prime}\right) & =3\chi_{ijkl}^{\left(3\right)}\left(\mathbf{r};-\omega,\omega,\omega^{\prime},-\omega^{\prime}\right)\label{eq:N.neq.single}\\
 & \times\left[b_{\mathrm{env}}\left(\omega^{\prime}\right)-b_{\mathrm{obj}}\left(\omega^{\prime}\right)\right]\nonumber \\
 & \times\left(\tilde{\mathbb{G}}\mathrm{Im}\left[-\mathbb{G}_{0}^{-1}\right]\tilde{\mathbb{G}}^{*}\right)\left(\mathbf{r},\mathbf{r};\omega^{\prime}\right)_{kl}.\nonumber 
\end{align}
We see that if there is only a single body in vacuum, the effective 
dielectric function depends on the temperature of the environment, 
in stark contrast to linear materials

\subsection{Heat radiation and transfer}

In Appendix \ref{sec:Heat-radiation-appendix}, we show that the net
heat radiated from a body can be written in terms of the fluctuation correlations
[see Eq.~\eqref{eq:Heat.general}], starting from the Poynting vector. By using the result obtained
in Eq.~\eqref{eq:ee.FF}, the net heat (including incoming and outgoing radiation)
from object $n$ in the presence of $N-1$ other objects can be written
as (derivation in Appendix \ref{sec:Heat-radiation-appendix})
\begin{align}
H_{n}= & \frac{1}{\mu_{0}}\sum_{m=0}^{N}\int\frac{d\omega}{2\pi}\frac{1}{\omega}\left[b_{n}\left(\omega\right)-b_{m}\left(\omega\right)\right]\nonumber \\
 & \times\mathrm{Tr}\left(\mathrm{Im}\left[\tilde{\mathbb{V}}_{m}\right]\mathrm{Im}\left[\tilde{\mathbb{G}}\tilde{\mathbb{V}}_{n}\tilde{\mathbb{G}}^{*}\right]\right),\label{eq:Heat.final.t}
\end{align}
We were not able to show that $\tilde{\mathbb{V}}$ and therefore $\tilde{\mathbb{G}}$ are generally symmetric (implying micro-reversibility \cite{Eckhardt1984}) in the considered non-equilibrium situation. This is why Eq.~\eqref{eq:Heat.final.t} is not symmetric in indices $n$ and $m$. If $\tilde{\mathbb{V}}$ is symmetric, the slightly simpler Eq.~\eqref{eq:Heat.final.symmetric-1} follows (see Appendix \ref{sec:Heat-radiation-appendix}), which is then symmetric in $n$ and $m$ like the corresponding formula for equilibrium systems \cite{kruger2012trace,Rodriguez2012}.

Eq.~\eqref{eq:Heat.final.t} reiterates the statement that in order to calculate the heat radiation or transfer, all 
we need to know are the effective linear properties of the system -- the effective 
nonequilibrium dielectric function or linear response. Recall that 
in the nonlinear case the effective dielectric function depends on the geometry and 
temperature of the rest of the system in a nontrivial fashion as per 
Eqs.~\eqref{eq:epsilon.effective-noneq} and \eqref{eq:N.full-noneq}.

Eq. \eqref{eq:Heat.final.t}, apart from the mentioned issue about symmetries, is similar in form to trace formulas
obtained in Refs.~\cite{Rodriguez2012,kruger2012trace}. Ref.~\cite{kruger2012trace}     writes it in terms of the scattering or T-operators 
\begin{align}
\tilde{\mathbb{T}}=\mathbb{H}_{0}\tilde{\mathbb{G}}\tilde{\mathbb{V}},\label{eq:T}
\end{align}
where the tilde again denotes the physical linear response. More precisely, formula Eq.~\eqref{eq:Heat.final.symmetric-1} in Appendix \ref{sec:Heat-radiation-appendix}, the symmetric version of Eq.~\eqref{eq:Heat.final.t}, is equivalent to the expressions of Refs.~\cite{kruger2012trace,Muller17}, when reduced to the linear system. 

For a single body in vacuum (assuming symmetry of $\tilde{\mathbb{V}}$), the heat radiation takes the form which is reminiscent of the corresponding result for linear systems \cite{kruger2012trace}, 
\begin{align}
H= & \frac{1}{\mu_{0}}\int\frac{d\omega}{2\pi}\frac{1}{\omega}\left[b_{\mathrm{obj}}\left(\omega\right)-b_{\mathrm{env}}\left(\omega\right)\right]\\
 & \times\mathrm{Tr}\left(\mathrm{Im}\left[\mathbb{G}_{0}\right]\mathrm{Im}\tilde{\mathbb{T}}-\mathrm{Im}\left[\mathbb{G}_{0}\right]\tilde{\mathbb{T}}\mathrm{Im}\left[\mathbb{G}_{0}\right]\tilde{\mathbb{T}}^{*}\right).\nonumber 
\end{align}
This equation follows from Eq.~\eqref{eq:Heat.final.symmetric-1}, when reduced to a
single body and substituting the identities \eqref{eq:T} and  $\tilde{\mathbb{G}}=\mathbb{G}_{0}+\mathbb{G}_{0}\tilde{\mathbb{T}}\mathbb{G}_{0}$.

\subsection{Heat radiation of a semi-infinite plate: Kirchhoff's law and Planck's
law}

We proceed by computing the nonequilibrium part of the effective dielectric
function for a single plate using Eq.~\eqref{eq:N.neq.single}.
In order to simplify the following discussion, we consider again a
highly symmetric material with $\varepsilon_{ij}=\varepsilon\delta_{ij}$
and $\chi_{ijkl}^{\left(3\right)}=\chi^{\left(3\right)}\delta_{ij}\delta_{kl}$.
In that case the effective dielectric function is diagonal and we
obtain from Eq.~\eqref{eq:N.neq.single}
\begin{align}
&\tilde{\varepsilon}^\mathrm{}_{ij}\left(\mathbf{r};\omega\right)= \delta_{ij}
\left[\varepsilon\left(\mathbf{r};\omega\right)+\int\mathrm{d}\omega^{\prime}N\left(\mathbf{r};\omega,\omega^{\prime}\right)\right],\label{eq:effective.epsilon.noneq.simple}\\
&N^{\mathrm{neq}}\left(\mathbf{r}\in V_{n};\omega,\omega^{\prime}\right) = 3\chi^{\left(3\right)}\left(\mathbf{r};-\omega,\omega,\omega^{\prime},-\omega^{\prime}\right) \nonumber\\
 &\hspace{4em} \times\sum_{m=0}^{N}\left[b_{\mathrm{env}}\left(\omega^{\prime}\right)-b_{\mathrm{obj}}\left(\omega^{\prime}\right)\right]\nonumber \\
 &\hspace{4em} \times\left(\tilde{\mathbb{G}}\mathrm{Im}\left[-\mathbb{G}_{0}^{-1}\right]\tilde{\mathbb{G}}^{*}\right)\left(\mathbf{r},\mathbf{r};\omega^{\prime}\right)_{kk},\label{eq:N.neq.simple}
\end{align}
where $N=N^{\mathrm{eq}}+N^{\mathrm{neq}}$, with the equilibrium
part given in Eq.~\eqref{eq:N.eq.iso}. 

\begin{figure}
\includegraphics[width=1\columnwidth]{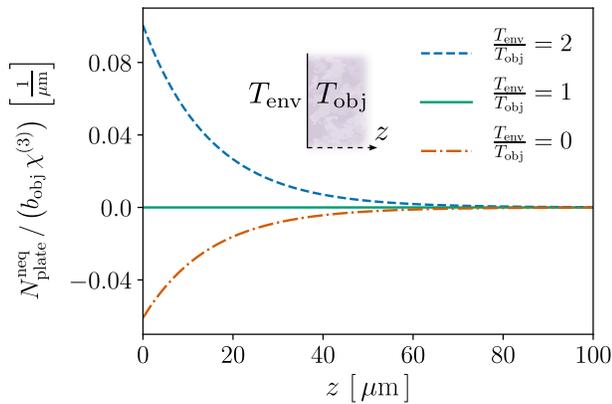} \caption{The nonequilibrium effective dielectric function inside a semi-infinite
plate, where $z$ is the distance from the surface. $\varepsilon=4+i$,
$\lambda_{0}^{\prime}=2\pi\frac{c}{\omega^{\prime}}=50\,\mathrm{\mu m}$,
and $T_{\mathrm{obj}}=\frac{\hbar\omega^{\prime}}{k_{\mathrm{B}}}\approx287\,\mathrm{K}$
were fixed while $T_{\mathrm{env}}$ was varied between zero and $2T_{\mathrm{obj}}$.
\label{fig:N.neq}}
\end{figure}

The numerical results (using again the Green's function for planar systems from 
Ref.~\cite{Johansson2011}) are shown in Fig.~\ref{fig:N.neq} for $\varepsilon=4+i$. 
The term $\tilde{\mathbb{G}}\mathrm{Im}\left[-\mathbb{G}_{0}^{-1}\right]\tilde{\mathbb{G}}^{*}$
 was evaluated using so-called environment dust \cite{Eckhardt1984,kruger2012trace},
which is finite, unlike $\mathrm{Im}\mathbb{G}\left(\mathbf{r},\mathbf{r}\right)$. We see that there is no interference pattern forming in $N_{\mathrm{plate}}^{\mathrm{neq}}$,
which is in contrast to the equilibrium dielectric function. Deep inside the material, i.e. for large $z$,  $N_{\mathrm{plate}}^{\mathrm{neq}}$  vanishes. For
non-absorbing media, $N_{\mathrm{plate}}^{\mathrm{neq}}$ is independent of $z$ (not shown).

Notably, the nonequilibrium contribution changes sign with $\Delta T=T_{\mathrm{obj}}-T_{\mathrm{env}}$, as can be seen from Eq.~\eqref{eq:N.neq.simple} since $b$ is 
monotonic in temperature. This implies
that with a non-zero imaginary part of $\chi^{\left(3\right)}$ (of
either sign), it is in principle possible to obtain $\mathrm{Im}\tilde{\varepsilon}<0$,
i.e., a medium with a negative absorption at the given frequency.
This means that a probing wave would experience gain in an otherwise
passive system due to interactions with the nonequilibrium fluctuations.
While there is no fundamental principle ruling this out (in contrast
to the equilibrium case), it remains to be seen whether materials with suitable 
combinations of $\chi^{(1)}$ and $\chi^{(3)}$ exist to display any such phenomenon.

More importantly, Figure \ref{fig:N.neq} shows that the effective
dielectric function (which determines heat radiation) of the plate
depends on the temperature of the environment, not just the plate itself.
This is a clear nonlinear effect (it is absent for linear materials), 
which can be measured experimentally. In the case of multiple bodies, the 
effective dielectric function also depends on the temperatures of other 
objects (in addition to their positions as in the equilibrium case).

This observation has implications for Kirchhoff's law of radiation.
It states that, in equilibrium, the absorptivity and emissivity of
a body are equal. It is thus a variant of the principle of detailed
balance. Although it strictly only holds in equilibrium, it has been
shown to be valid (and often used) for linear bodies out of equilibrium
as well (see e.g. the discussion in Ref.~\cite{kruger2012trace}).
The analysis in section \ref{sec:NeqR}, as well as Fig.~\ref{fig:N.neq},
shows an interesting variant of this law: Because the nonequilibrium
effective dielectric function can be used for computation of absorption
as well as emission coefficients, Kirchoff's law stays indeed valid
in the considered order of $\chi^{(3)}$. However, these coefficients
depend on the temperature of the environment. This means that the experiments
measuring the emission and absorption need to be performed in exactly the 
same conditions (same temperatures, surrounding bodies, etc.).

We noted in the previous subsection that we cannot prove the symmetry of the nonequilibrium potential $\tilde{\mathbb{V}}$ in general, mostly due to lack of symmetries of  $\chi^{(3)}_{ijkl}$. For the highly symmetric version used in Fig.~\ref{fig:N.neq}, $\chi_{ijkl}^{\left(3\right)}=\chi^{\left(3\right)}\delta_{ij}\delta_{kl}$, $\tilde{\mathbb{V}}$ is symmetric. If $\tilde{\mathbb{V}}$ may turn out to be non-symmetric in other cases, this would manifest a more dramatic breakdown of Kirchoff's law, because such non-symmetric non-equilibrium $\tilde{\mathbb{V}}$ would explicitly break micro-reversibility. This has to be investigated in the future.


Noting the change of the dielectric function in Fig.~\ref{fig:N.neq},
one may ask whether the plate in that figure may radiate stronger
than a black body. This question is immediately answered from
the observation that the radiation of the plate is given by known
formulas (see e.g. \cite{kruger2012trace}), where the dielectric function
$\varepsilon$ needs to be replaced by the effective one of Fig.~\ref{fig:N.neq}.
While explicit computation of the corresponding Fresnel coefficients
for spatially varying $\varepsilon$ may be challenging, a general
statement is nevertheless possible. The radiation of a planar
surface, irrespective of the values of $\varepsilon(\omega)$, is
positive and bound by the radiation of a black body. Therefore, for the radiation linear in $\chi^{(3)}$, we have
\begin{equation}
0\leq d\omega \frac{{\cal H}\left(\omega\right)}{A} \leq 
d\omega \frac{\hbar \omega^3}{4 \pi^2 c^2} 
\left[ \exp\left(\frac{\hbar\omega}{k_\mathrm{B}T}\right)-1\right]^{-1},\label{eq:Pl}
\end{equation}
where the radiation of the plate per surface area $A$ is $H/A=\int_0^\infty d\omega {\cal H}(\omega)/A$.
The radiation of a planar surface thus obeys the fundamental bounds implied by Planck's law. We note, however, that Eq.~\eqref{eq:Pl} relies on the symmetry of $\tilde{\mathbb{V}}$. Again, the possiblity of non-symmetric $\tilde{\mathbb{V}}$ out of equilibrium must be investigated in the future.

\subsection{Radiation of a sphere: negative radiation}

We now turn to the radiation of a nanosphere. We start by evaluating
the isotropic effective dielectric function in Eq.~\eqref{eq:effective.epsilon.noneq.simple} using also the simplification $\chi_{ijkl}^{\left(3\right)}=\chi^{\left(3\right)}\delta_{ij}\delta_{kl}$ 
as in Sec.~\ref{subsec:The-effective-dielectric}.
In the limit where the radius is much smaller than the thermal wavelength
$\lambda_{T}=\frac{\hbar c}{k_{\mathrm{B}}T}$ and the skin depth
$\delta=\frac{1}{\mathrm{Im}\sqrt{\varepsilon\mu}}\frac{c}{\omega}$,
the Green's function with one point outside and one point inside the
sphere is given by $\mathbb{G}=\frac{3}{\varepsilon+2}\mathbb{G}_{0}$.
Using $\mathrm{Im}\mathbb{G}_{0}\left(\mathbf{r},\mathbf{r};\omega\right)_{ij}=\frac{1}{6\pi}\frac{\omega}{c}\delta_{ij}$,
we have from Eq.~\eqref{eq:N.neq.simple}, 
\begin{align}
N_{\mathrm{sphere}}^{\mathrm{neq}}\left(\omega,\omega^{\prime}\right) & =-\frac{3}{2\pi}\frac{\omega^{\prime}}{c}\chi^{\left(3\right)}\left(-\omega,\omega,\omega^{\prime},-\omega^{\prime}\right)\label{eq:N.sphere}\\
 & \times\left|\frac{3}{\varepsilon\left(\omega^{\prime}\right)+2}\right|^{2}\left[b_{\mathrm{obj}}\left(\omega^{\prime}\right)-b_{\mathrm{env}}\left(\omega^{\prime}\right)\right].\nonumber 
\end{align}
This function is spatially constant inside the (point-like) sphere.
We may now use this dielectric function to compute the effective version
of the polarizability 
\begin{align}
\alpha\equiv\frac{\varepsilon-1}{\varepsilon+2}R^{3},\label{eq:alpha}
\end{align}
which governs the radiation of small spheres \cite{Tsang2000,bohren2008}.
By substituting the effective dielectric function into Eq.~\eqref{eq:alpha}
and expanding in $N_{\mathrm{sphere}}^{\mathrm{neq}}$, we obtain
\begin{equation}
\tilde{\alpha}\left(\omega\right)=\tilde{\alpha}^{\mathrm{eq}}\left[1+\frac{3}{\left(\varepsilon-1\right)\left(\varepsilon+2\right)}\int d\omega^{\prime}N_{\mathrm{sphere}}^{\text{neq}}\left(\omega,\omega^{\prime}\right)\right],\label{eq:alpha.effective}
\end{equation}
where $\tilde{\alpha}^{\mathrm{eq}}$ is the (effective) polarizability
in equilibrium. The radiation 
of a sphere is then given by  
\begin{equation}
H=4\frac{\varepsilon_{0}}{\pi^{2}c}\int d\omega\omega^{2}\left[b_{\mathrm{obj}}\left(\omega\right)-b_{\mathrm{env}}\left(\omega\right)\right]\mathrm{Im}\tilde{\alpha}\left(\omega\right).\label{eq:H.nanosphere}
\end{equation}
$\mathrm{Im}\tilde{\alpha}$, which is manifestly positive in equilibrium, may in principle be negative in the considered non-equilibrium situation, for suitable regimes regarding the sign of $(T_{\mathrm{env}}-T_{\mathrm{obj}})$
as well as $\mathrm{Im}\chi^{\left(3\right)}\left(-\omega,\omega\right)$.
As mentioned before, we see no fundamental reason that forbids
such occurrence, and it will be interesting to see whether it can exist
in practice.

An instructive extreme case to consider is $\mathrm{Im}\tilde{\varepsilon}^{\mathrm{eq}}=\mathrm{Im}\tilde{\alpha}^{\mathrm{eq}}=0$. This is a particle that does not absorb or emit energy in equilibrium, so that any absorption is only due to the finite $\textrm{Im}\chi^{(3)}$.
Using Eqs.~\eqref{eq:N.sphere}, \eqref{eq:alpha.effective}, and \eqref{eq:H.nanosphere}, we then arrive at
\begin{align}
H & =-54\frac{\varepsilon_{0}}{\pi^{3}c^{3}}\int d\omega\int d\omega^{\prime}\omega^{2}\omega^{\prime}\nonumber\mathrm{Im}\chi^{\left(3\right)}\left(-\omega,\omega,\omega^{\prime},-\omega^{\prime}\right)  \\
 &\frac{\left(b_{\mathrm{obj}}\left(\omega\right)-b_{\mathrm{env}}\left(\omega\right)\right)\left(b_{\mathrm{obj}}\left(\omega^{\prime}\right)-b_{\mathrm{env}}\left(\omega^{\prime}\right)\right) }
{\left(\varepsilon\left(\omega\right)+2\right)^2\left(\varepsilon\left(\omega^{\prime}\right)+2\right)^{2}} \label{eq:H.detailed}.
\end{align}
The first observation regarding Eq.~\eqref{eq:H.detailed} is that the heat radiation of the sphere remains {\it unchanged} if the temperatures of the object and the environment are interchanged. Considering specifically $\textrm{Im}\chi^{\left(3\right)}<0$, which is a typical observed case (see e.g. Ref~\cite{Hamanaka2003} for metal-infused glasses),  
Eq.~\eqref{eq:H.detailed} yields $H>0$. This means energy flowing away from the sphere for any combination of temperatures. For $T_{\mathrm{env}}>T_{\mathrm{obj}}$, this corresponds to a flow of enery from a cold sphere to a hot environment. While this cannot be ruled out for a particular frequency, from thermodynamic
considerations we expect that the total heat (after integration over all
frequencies) flows from the hotter to the colder body.

\section{Summary}
In stochastic nonlinear optical systems, fluctuating fields and induced fields couple, which gives rise to a variety of phenomena which cannot be observed in purely linear systems. We show that fluctuation effects, such as the Casimir effect or heat radiation, can be described via known formulas, however using an effective dielectric function as input. This dielectric function depends on the shape of the objects, their relative position, and also on the temperatures of all objects in the system.  

The divergence of the electromagnetic Green's function at coinciding points prevents a straight computation of the effective dielectric function on a macroscopic level for absorbing materials. It is nevertheless possible to circumvent this issue by considering \emph{measurable} quantities. Using this principle, the dependence of the dielectric function on the distance between the objects is accessible theoretically. This is also true for the dependence of the dielectric function of one object on the temperatures of the other objects in a non-equilibrium scenario.  

In addition to effects in equilibrium, we saw profound and thought-provoking implications in the case where temperatures of objects (and the environment) are different. We discussed the applicability of  Kirchoff's law of radiation as well as the fundamental bounds of radiation of a planar surface. For a nano-sphere out of equilibrium, we found that the spectral emission can surprisingly be negative in certain cases. 

Overall, we saw that both equilibrium and nonequilibrium phenomena are intricately affected by nonlinear optical properties. By using the fluctuational electrodynamics framework, these results are also applicable for any geometry or materials. In the future it may also be generalized to nonzero external fields, possibly allowing for even more control over the effect of the nonlinearities.

\section{Acknowledgments}

This work was supported by Deutsche Forschungsgemeinschaft (DFG) Grant No. KR 3844/2-1 and MIT-Germany Seed Fund Grant No. 2746830. 

\appendix

\section{Heat radiation and transfer from fluctuational electrodynamics\label{sec:Heat-radiation-appendix}}

The total energy transmitted across a surface $\Sigma_{n}$ surrounding object $n$ is given by 
\begin{equation}
H_{n}=\oint_{\Sigma_{n}}da\left\langle \mathbf{S}\right\rangle \cdot\mathbf{n},\label{eq:Hn.da}
\end{equation}
where $\left\langle \mathbf{S}\right\rangle =\left\langle \mathbf{E}\times\mathbf{H}\right\rangle$ is the time-average of the Poynting vector and $\mathbf{n}$ is a normal vector on $\Sigma_{n}$. The force on the object (called Casimir force in equilibrium) can be obtained from the same expression with the Poynting vector replaced by the Maxwell stress tensor $\sigma=\varepsilon_{0}\left(\left\langle \mathbf{E}\otimes\mathbf{E}\right\rangle -\frac{1}{2}\mathbf{E}^{2}\right)+\frac{1}{\mu_{0}}\left(\left\langle \mathbf{B}\otimes\mathbf{B}\right\rangle -\frac{1}{2}B^{2}\right)$.

For stationary systems, the correlator $\left\langle \mathbf{E}\left(t\right)\otimes\mathbf{H}\left(t^{\prime}\right)\right\rangle $ depends only on time differences. We can therefore define a spectral density of the expectation value as 
\begin{equation}
\left\langle \mathbf{E}\left(t\right)\otimes\mathbf{H}\left(t^{\prime}\right)\right\rangle =\int\frac{d\omega}{2\pi}e^{i\omega(t-t^{\prime})}\left\langle \mathbf{E}\otimes\mathbf{H}^{*}\right\rangle _{\omega},
\end{equation}
where the integration is over positive and negative frequencies. Since $\left\langle \mathbf{E}\left(t\right)\otimes\mathbf{H}\left(t^{\prime}\right)\right\rangle $ is a real quantity, the real (imaginary) part of the spectrum is an even (odd) function of the frequency. Therefore only the real part remains in the Poynting vector 
\begin{equation}
\left\langle \mathbf{S}\right\rangle =\int\frac{d\omega}{2\pi}\mathrm{Re}\left\langle \mathbf{E}\times\mathbf{H}^{*}\right\rangle _{\omega}.
\end{equation}

Using the divergence theorem, we can rewrite Eq. \eqref{eq:Hn.da} as 
\begin{align}
H_{n} & =\int_{V_{n}}dV\left\langle \nabla\cdot\mathbf{S}\right\rangle \\
 & =\int\frac{d\omega}{2\pi}\int_{V_{n}}dV\mathrm{Re}\left\langle \nabla\cdot\left(\mathbf{E}_{\omega}\times\mathbf{H}_{\omega}^{*}\right)\right\rangle .\nonumber 
\end{align}
With the Maxwell-Faraday equation $-i\omega\mathbf{B}_{\omega}^{*} =\nabla\times\mathbf{E}_{\omega}^{*}$,
we can write for nonmagnetic materials ($\mu=1$)
\begin{equation}
H_{n}=\frac{1}{\mu_{0}}\int\frac{d\omega}{2\pi}\frac{1}{\omega}\int_{V_{n}}dV\mathrm{Im}\left\langle \mathbf{E}\cdot\left(\nabla\times\nabla\times\mathbf{E}\right)^{*}\right\rangle _{\omega}.
\end{equation}
By using the symmetric operator $\mathbb{G}_{0}^{-1}=\mathbb{H}_{0}=\nabla\times\nabla\times-\frac{\omega^{2}}{c^{2}}\mathbb{I}$, we get 
\begin{equation}
H_{n}=\frac{1}{\mu_{0}}\int\frac{d\omega}{2\pi}\frac{1}{\omega}\mathrm{Tr}_{n}\mathrm{Im}\left[\left\langle \mathbf{E}\otimes\mathbf{E}^{*}\right\rangle _{\omega}\mathbb{G}_{0}^{-1}\right],\label{eq:Heat.general}
\end{equation}
where $\mathrm{Tr}_{n}$ denotes a trace, which is restricted to volume $V_n$. Here we see the imaginary part of the electric field correlator. In equilibrium it is zero [see Eq. \eqref{eq:FDT}] and no energy is transferred, as expected. Out of equilibrium, however, the correlator in Eq. \eqref{eq:ee.FF} can obtain a nonzero imaginary part which results in a net heat transfer from or to the body. 

Substituting it into Eq. \eqref{eq:Heat.general} and subtracting the case where all temperatures are equal to $T_{n}$ gives
\begin{align}
H_{n}= & \frac{1}{\mu_{0}}\sum_{m=0}^{N}\int\frac{d\omega}{2\pi}\frac{1}{\omega}\left[b_{m}\left(\omega\right)-b_{n}\left(\omega\right)\right]\nonumber \\
 & \times\mathrm{Im}\mathrm{Tr}_{n}\left(\tilde{\mathbb{G}}\mathrm{Im}\left[\tilde{\mathbb{V}}_{m}\right]\tilde{\mathbb{G}}^{*}\mathbb{G}_{0}^{-1}\right).
\end{align}
Using the identity $\tilde{\mathbb{G}}=\left(\mathbb{I}+\tilde{\mathbb{G}}\tilde{\mathbb{V}}\right)\mathbb{G}_{0}$, the free Green's functions are canceled and we obtain the heat radiation as 
\begin{align}
H_{n}= & \frac{1}{\mu_{0}}\sum_{m=0}^{N}\int\frac{d\omega}{2\pi}\frac{1}{\omega}\left[b_{n}\left(\omega\right)-b_{m}\left(\omega\right)\right]\nonumber \\
 & \times\mathrm{Tr}\left(\mathrm{Im}\left[\tilde{\mathbb{V}}_{m}\right]\mathrm{Im}\left[\tilde{\mathbb{G}}\tilde{\mathbb{V}}_{n}\tilde{\mathbb{G}}^{*}\right]\right),\label{eq:Heat.final}
\end{align}
where now the full trace appears, which allows for cyclic rearrangement of the operators. Note  none of terms with $T_{m}=T_{n}$ (including $m=n$) contribute to heat radiation. 

Furthermore, if $\tilde{\mathbb{V}}$ and therefore $\tilde{\mathbb{G}}$ are symmetric (implying micro-reversibility \cite{Eckhardt1984}), then we can further simplify Eq. \eqref{eq:Heat.final} as
\begin{align}
H_{n}= & \frac{1}{\mu_{0}}\sum_{m=0}^{N}\int\frac{d\omega}{2\pi}\frac{1}{\omega}\left[b_{n}\left(\omega\right)-b_{m}\left(\omega\right)\right]\nonumber \\
 & \times\mathrm{Tr}\left(\mathrm{Im}\left[\tilde{\mathbb{V}}_{m}\right]\tilde{\mathbb{G}}\mathrm{Im}\left[\tilde{\mathbb{V}}_{n}\right]\tilde{\mathbb{G}}^{*}\right).\label{eq:Heat.final.symmetric-1}
\end{align}
This is the final form of for the net heat radiation from object $n$ to the environment and other objects.

%

\end{document}